
\null
\nopagenumbers
\vskip -20pt
\centerline{\subsection Theoretical Physics Institute}
\centerline{\subsection University of Minnesota}
\vskip 1.2truecm
\line {\hfill TPI-MINN-92/59-T}
\line {\hfill October 1992}
\vskip 1.2truecm
\centerline {\subsection Fermi-Dirac Corrections to the Relic Abundances}
\vskip 1.5truecm
\centerline {A.\ D.\ Dolgov$^1$ and K.\ Kainulainen$^{2}$}
\vskip 0.4truecm
\centerline {$^1$ Department of Physics,
University of Michigan, Ann Arbor, MI 48109}
\centerline {and}
\centerline {ITEP, Moscow 117259, Russia}
\centerline {$^2$Theoretical Physics Institute, University of Minneapolis}
\centerline {Minneapolis MN 55455}
\\

  \magnification=1200
  \hoffset 0.3truecm  
  \hsize=15.6truecm   
  \vsize=23.0truecm
  \topskip=20pt            
  \fontdimen1\tenrm=0.0pt  
  \fontdimen2\tenrm=4.0pt  
  \fontdimen3\tenrm=7.0pt  
  \fontdimen4\tenrm=1.6pt  
  \fontdimen5\tenrm=4.3pt  
  \fontdimen6\tenrm=10.0pt 
  \fontdimen7\tenrm=2.0pt  
  \baselineskip=12.0pt plus 1.0pt minus 0.5pt  
  \lineskip=1pt plus 0pt minus 0pt             
  \lineskiplimit=1pt                           
  \parskip=2.5pt plus 5.0pt minus 0.5pt
  \parindent=15.0pt
  \baselineskip=0.7truecm
\font\section=cmbx10 scaled\magstep2
\font\subsection=cmbx10 scaled\magstep1

\def\\{\vskip 0mm}

\def\jump{\vskip 0.5truecm}

\def\ssstyle{\scriptscriptstyle}
%
\def\parbarit#1{#1\kern-0.8em\raise1.5ex\hbox{$\ssstyle(-)$}}
\def\parbar#1{#1\kern-1.0em\raise1.5ex\hbox{$\scriptscriptstyle(-)$}}
\def\Parbar#1{#1\kern-1.0em\raise2.1ex\hbox{$\scriptscriptstyle(-)$}}
\def\Slash#1{#1\kern-0.5em\raise.05ex\hbox{/}}
\def\slash#1{#1\kern-0.45em\raise.05ex\hbox{{$\scriptstyle /$}}}

\def\oo{\slash o}

\def\lsim{\;\raise0.3ex\hbox{$<$\kern-0.75em\raise-1.1ex\hbox{$\sim$}}\;}
\def\gsim{\;\raise0.3ex\hbox{$>$\kern-0.75em\raise-1.1ex\hbox{$\sim$}}\;}
\def\ref#1{[{#1}]}

\def\vs{\it vs.}
%

%

%









\newcount\refnumber
\newcount\temp
\newcount\test
\newcount\tempone
\newcount\temptwo
\newcount\tempthr
\newcount\tempfor
\newcount\tempfiv
\newcount\testone
\newcount\testtwo
\newcount\testthr
\newcount\testfor
\newcount\testfiv
\newcount\itemnumber
\newcount\totalnumber
\refnumber=0
\itemnumber=0
\def\initreference#1{\totalnumber=#1
                 \advance \totalnumber by 1
                 \loop \advance \itemnumber by 1
                       \ifnum\itemnumber<\totalnumber
                        \temp=100 \advance\temp by \itemnumber
                        \count\temp=0 \repeat}

\def\ref#1{\temp=100 \advance\temp by #1
   \ifnum\count\temp=0
    \advance\refnumber by 1  \count\temp=\refnumber \fi
   \ [\the\count\temp]}

\def\reftwo#1#2{\tempone=100 \advance\tempone by #1
   \ifnum\count\tempone=0
   \advance\refnumber by 1  \count\tempone=\refnumber \fi
   \temptwo=100 \advance\temptwo by #2
   \ifnum\count\temptwo=0
   \advance\refnumber by 1  \count\temptwo=\refnumber \fi
 \testone=\count\tempone \testtwo=\count\temptwo
 \sorttwo\testone\testtwo
     \ [\the\testone,\the\testtwo]}       

\def\refthree#1#2#3{\tempone=100 \advance\tempone by #1
   \ifnum\count\tempone=0
    \advance\refnumber by 1  \count\tempone=\refnumber \fi
    \temptwo=100 \advance\temptwo by #2
   \ifnum\count\temptwo=0
    \advance\refnumber by 1  \count\temptwo=\refnumber \fi
    \tempthr=100 \advance\tempthr by #3
   \ifnum\count\tempthr=0
    \advance\refnumber by 1  \count\tempthr=\refnumber \fi
 \testone=\count\tempone \testtwo=\count\temptwo \testthr=\count\tempthr
 \sortthree\testone\testtwo\testthr
   \test=\testthr  \advance\test by -2
 \ifnum\test=\testone    \test=\testtwo  \advance\test by -1
    \ifnum\test=\testone   
    \ [\the\testone--\the\testthr]\fi \advance\temptwo by 1
  \else
     \ [\the\testone,\the\testtwo,\the\testthr]    
 \fi}

\def\reffour#1#2#3#4{\tempone=100 \advance\tempone by #1
   \ifnum\count\tempone=0
    \advance\refnumber by 1  \count\tempone=\refnumber \fi
    \temptwo=100 \advance\temptwo by #2
   \ifnum\count\temptwo=0
    \advance\refnumber by 1  \count\temptwo=\refnumber \fi
    \tempthr=100 \advance\tempthr by #3
   \ifnum\count\tempthr=0
    \advance\refnumber by 1  \count\tempthr=\refnumber \fi
    \tempfor=100 \advance\tempfor by #4
   \ifnum\count\tempfor=0
    \advance\refnumber by 1  \count\tempfor=\refnumber \fi
 \testone=\count\tempone \testtwo=\count\temptwo \testthr=\count\tempthr
 \testfor=\count\tempfor
 \sortfour\testone\testtwo\testthr\testfor
   \test=\testthr \advance\test by -1
   \ifnum\testtwo=\test   \test=\testtwo \advance\test by -1
    \ifnum\testone=\test  \test=\testfor \advance\test by -3
     \ifnum\testone=\test \ [\the\testone--\the\testfor]
     \else \ [\the\testone--\the\testthr,\the\testfor]
     \fi
    \else  \test=\testfor \advance\test by -1
     \ifnum\testthr=\test \ [\the\testone,\the\testtwo--\the\testfor]
     \else\ [\the\testone,\the\testtwo,\the\testthr,\the\testfor]
     \fi
    \fi
   \else \ [\the\testone,\the\testtwo,\the\testthr,\the\testfor]
   \fi}

\def\reffive#1#2#3#4#5{\tempone=100 \advance\tempone by #1
   \ifnum\count\tempone=0
    \advance\refnumber by 1  \count\tempone=\refnumber \fi
    \temptwo=100 \advance\temptwo by #2
   \ifnum\count\temptwo=0
    \advance\refnumber by 1  \count\temptwo=\refnumber \fi
    \tempthr=100 \advance\tempthr by #3
   \ifnum\count\tempthr=0
    \advance\refnumber by 1  \count\tempthr=\refnumber \fi
    \tempfor=100 \advance\tempfor by #4
   \ifnum\count\tempfor=0
    \advance\refnumber by 1  \count\tempfor=\refnumber \fi
    \tempfiv=100 \advance\tempfiv by #5
   \ifnum\count\tempfiv=0
    \advance\refnumber by 1  \count\tempfiv=\refnumber \fi
 \testone=\count\tempone \testtwo=\count\temptwo \testthr=\count\tempthr
 \testfor=\count\tempfor \testfiv=\count\tempfiv
 \sortfive\testone\testtwo\testthr\testfor\testfiv
  \test=\testthr \advance\test by -1
  \ifnum\testtwo=\test   \test=\testtwo \advance\test by -1
   \ifnum\testone=\test  \test=\testfor \advance\test by -3
    \ifnum\testone=\test \test=\testfiv \advance\test by -4
     \ifnum\testone=\test\ [\the\testone--\the\testfiv]
     \else\ [\the\testone--\the\testfor,\the\testfiv]
     \fi
    \else \ [\the\testone--\the\testthr,\the\testfor,\the\testfiv]
    \fi
   \else  \test=\testfor \advance\test by -1
    \ifnum\testthr=\test \test=\testfiv \advance\test by -2
     \ifnum\testthr=\test \ [\the\testone,\the\testtwo--\the\testfiv]
     \else \ [\the\testone,\the\testtwo--\the\testfor,\the\testfiv]
     \fi
    \else\ [\the\testone,\the\testtwo,\the\testthr,\the\testfor,\the\testfiv]
    \fi
   \fi
  \else \test=\testfor \advance\test by -1
   \ifnum\testthr=\test \test=\testfiv \advance\test by -2
    \ifnum\testthr=\test\
[\the\testone,\the\testtwo,\the\testthr--\the\testfiv]
    \else\ [\the\testone,\the\testtwo,\the\testthr,\the\testfor,\the\testfiv]
    \fi
   \else\ [\the\testone,\the\testtwo,\the\testthr,\the\testfor,\the\testfiv]
   \fi
  \fi}

\def\refitem#1#2{\temp=#1 \advance \temp by 100 \setbox\count\temp=\hbox{#2}}

\def\sortfive#1#2#3#4#5{\sortfour#1#2#3#4\relax
   \ifnum#5<#4\relax \test=#5\relax #5=#4\relax
     \ifnum\test<#3\relax #4=#3\relax
       \ifnum\test<#2\relax #3=#2\relax
         \ifnum\test<#1\relax  #2=#1\relax  #1=\test
         \else #2=\test \fi
       \else #3=\test \fi
     \else #4=\test \fi \fi}

\def\sortfour#1#2#3#4{\sortthree#1#2#3\relax
    \ifnum#4<#3\relax \test=#4\relax #4=#3\relax
       \ifnum\test<#2\relax #3=#2\relax
          \ifnum\test<#1\relax #2=#1\relax #1=\test
          \else #2=\test \fi
       \else #3=\test \fi \fi}

\def\sortthree#1#2#3{\sorttwo#1#2\relax
       \ifnum#3<#2\relax \test=#3\relax #3=#2\relax
          \ifnum\test<#1\relax #2=#1\relax #1=\test
          \else #2=\test \fi \fi}

\def\sorttwo#1#2{\ifnum#2<#1\relax \test=#2\relax #2=#1\relax #1=\test \fi}


\def\setref#1{\temp=100 \advance\temp by #1
   \ifnum\count\temp=0
    \advance\refnumber by 1  \count\temp=\refnumber \fi}

\def\printreference{\totalnumber=\refnumber
           \advance\totalnumber by 1
           \itemnumber=0
           \loop \advance\itemnumber by 1  
                 \ifnum\itemnumber<\totalnumber
                 \item{[\the\itemnumber]} \unhbox\itemnumber \repeat}


\def\dr{{\rm d}}
\def\ir{{\rm i}}
\def\nr{{\rm n}}
\def\cP{{\cal P}}
\def\vs{\langle v_{\rm M\oo l}\sigma^{(\nr )} \rangle}
\def\fnstyle{\baselineskip8pt \font \bx cmr8 \bx}

 \\

\vskip 2.0truecm
\centerline {\bf Abstract}
\vskip 1truecm
\hskip -15pt
We derive an equation for the evolution of the number density of a massive
particle species in the early Universe, which correctly accounts for the
Fermi-Dirac (FD) statistics.  The FD-corrections are sizable and potentially
important if the decoupling from the thermal equilibrium takes place at
temperatures of the order of, or less than the mass of the particle.
This is the case e.g.\ for a few MeV tau neutrino with the
ordinary weak interactions.
\vfill\eject
\end
\jump
\centerline {\subsection 1. Introduction}
\jump
Calculation of the evolution of the number densities of new hypothetical
massive particles has become
a very important part of the study of the early universe. A lot of work
has been devoted to the calculation of relic abundances of various cold
dark matter candidates in the present day universe
\reffive {1}{70}{100}{3}{101}.  It is
usually correct to assume that the particles remain in thermal equilibrium to
temperatures much below their mass, so that around the decoupling temperature
$T_\dr$ it is adequate to approximate the Fermi-Dirac (FD) statistics with the
Maxwell-Boltzmann (MB) statistics.  One can then show that in these
temperatures
($T \ll m$) the number density $n$ is governed by the equation
\reffive{71}{1}{2}{4}{5}
$$
{\dr n \over \dr t} + 3Hn  =  \langle v_{\rm M\oo l}\sigma \rangle
(n_{\rm eq}^2 - n^2),
\eqno(1)$$
where $n_{\rm eq}$ is the equilibrium density
and $H = \left({4\pi^3\over 45}g(T){T^4\over M_{\rm Pl}^2} \right)^{1/2}$ is
the Hubble expansion factor where $M_{\rm Pl}$ is the Planck Mass and the
function $g(T)$ is related to the total energy density $\rho$ by $\rho (T)
\equiv {\pi^2\over 30} g(T) T^4$. $\langle v_{\rm M\oo l}\sigma \rangle$ is a
thermally averaged cross section to be defined precisely below; here it
suffices to note that it is independent of $n$.  This separation of the R.H.S.\
of the equation (1) into a power function of $n$ multiplied by a $t$-dependent
factor can only be obtained in the Maxwell-Boltzmann approximation.

If the decoupling temperature is of the order of the mass, it is not correct to
use the MB-statistics and hence the equation (1) is no more adequate.  One
potentially interesting physical candidate for which $T_\dr \sim m$ is a few
MeV
tau neutrino.  A tau neutrino in this mass range would have a significant
effect on the nucleosynthesis \ref{6}.

In general it is not possible to write down an equation for the number density
$n$ directly, but as we shall show below, a tractable equation can be found
for the so called pseudo-chemical potential $z(t)$ \reftwo{72}{2}, that is of
the functional form
$$
{\dr z\over \dr t} = F(z(t),t).
\eqno(2)$$
Knowledge of $z(t)$ allows the calculation of all thermodynamical quantities
as simple integrals over momentum at each time $t$.

In section 2 we will derive the explicit form of the equation (2). In section 3
we will derive the MB-limit for that equation and compare our result to the
the literature. In section 4 we solve our equations numerically
in a simple example and compare the solution of the equation (2)
with the correct statistics to the numerical as well as to one often
used approximative solution of the equation (1). Section 5 contains
our conclusions and discussion.
\jump
\centerline {\subsection 2. The evolution equation}
\jump
We will derive the evolution equation for fermions (called neutrinos in what
follows) under the standard
assumption made in the literature
that all the helicity states are equally
populated \setref{8}
\footnote{*}{{\fnstyle Note that this is not necessarily true
in a realistic case, where the interactions may be chiral.
This problem is considered
elsewhere \ref{8}.}}.
Our starting point is then the Boltzmann equation for the scalar
distribution function $f(p,t)$ in the flat Friedmann-Robertson-Walker space
time \reftwo{2}{7}
$$
E(\partial_t + pH\partial_p)f(p,t) = C_{\rm E}(p,t) + C_{\rm I}(p,t),
\eqno(3)$$
where $E = (p^2 + m^2)^{1/2}$ and $C_{\rm E}(p,t)$ and $C_{\rm I}(p,t)$ are the
elastic and inelastic collision integrals respectively.  Elastic collisions are
responsible for maintaining the kinetic equilibrium, but the exact form of the
elastic collision integral is not important for our present purposes.  The
inelastic collision integral on the other hand is given by
$$
\eqalign{
C_{\rm I}(p_1,t) = {1\over 2} & \sum_\nr \int \prod_{\ir =2}^4
{\dr^3p_\ir \over (2\pi )^32E_\ir } (2\pi )^4\delta^4(p_1+p_2-p_3-p_4)\cr
&\sum_{s_\ir ,\ir =2}^4 \lbrace  \phantom{n}
{\mid {\cal M}_{34\rightarrow 12}^{(\nr )} \mid }^2
f_\nr (p_3,t){\bar f}_\nr (p_4,t)\left[ 1-f(p_1,t) \right]
\left[ 1-{\bar f}(p_2,t) \right] \cr
&\phantom{\sum n} - {\mid {\cal M}_{12\rightarrow 34}^{(\nr )} \mid }^2
f(p_1,t){\bar f}(p_2,t)\left[ 1-f_\nr (p_3,t) \right]
\left[ 1-{\bar f}_\nr (p_4,t) \right]  \phantom{l} \rbrace,  \cr
}\eqno(4)$$
where the sum over n includes all the particles in the system that the
neutrinos interact with and the sum over $s_\ir$ goes over the spins.
Note the normalization factor $1\over 2$ in front of the integral,
which is
sometimes forgotten in the literature.
The overbars above the distribution functions refer to antiparticles.  In what
follows we will assume vanishing chemical potentials so that
for all the particles $\bar f = f$.

In the usual approach leading to the equation (1), the following assumptions
are made:

\indent (i) The neutrinos are in kinetic equilibrium. \hfill\break
\indent (ii) Particles involved in the sum over n in the equation
(4) are in complete \hfill\break
\indent $\phantom {\rm (ii)}$ thermal equilibrium. \hfill\break
\indent (iii) MB-statistics is adequate for neutrinos.

Here we will also make the assumptions (i) and (ii), but we will relax the
assumption (iii).  Due to (i) and (ii) we can write the distribution functions
as
$$
f(p,t) \rightarrow f(p,z) \equiv (e^{\beta E + z}+1)^{-1}
\eqno(5{\rm a})$$
$$
f_\nr (p,t) \rightarrow f_\nr (p) \equiv (e^{\beta E_\nr}+1)^{-1},
\eqno(5{\rm b})$$
where $\beta \equiv 1/T$ and $z(t)$ is the pseudo-chemical potential
\reftwo{72}{2}.
It should be noted that, in contrast to the usual chemical potentials
in equilibrium,
$z(t)$ appears with the same sign in both the distributions for particles
and antiparticles. Inserting (5a) and (5b) into (4) and by using
unitarity and CPT or
the principle of detailed balance one can integrate the R.H.S.\ of the
equation (3) and get
$$\eqalign{
S_{\rm I}(z) &\equiv \sum_{\rm spin} \int
{\dr^3p\over (2\pi )^3} {1\over E}C_{\rm I}(p,t)\cr
&= (e^{2z} - 1) \sum_\nr \int \prod_{\ir =1}^4 {\dr^3p_\ir\over
(2\pi )^32E_\ir}(2\pi )^4\delta^4(p_1+p_2-p_3-p_4) \cr
&\phantom{kati} \times \sum_{s_\ir, \ir=1}^4
{\mid {\cal M}_{12\rightarrow 34}^{(\nr)} \mid}^2
f(p_1,z)f(p_2,z)\left[ 1-f_\nr(p_3) \right] \left[ 1-f_\nr(p_4)
\right].\cr
}\eqno(6)$$
Of course, similar integral over the elastic collision term $C_{\rm E}(p,t)$
vanishes under the conditions (i) and (ii).  Performing the same
operation to the L.H.S.\ of the equation (3) we find
$$
\sum_{\rm spin} \int {\dr^3p \over (2\pi )^3}(\partial_t -
pH\partial_p)f(p,t) = - A(z){\dr z \over \dr t} + B(z),
\eqno(7)$$
where the functions $A(z)$ and $B(z)$ are defined as
$$\eqalign{
A(z) &\equiv g\int {\dr^3p \over (2\pi )^3} f(p,z)^2 e^{\beta E+z} \cr
B(z) &\equiv g\int {\dr^3p \over (2\pi )^3} \left( (H+{1\over T}{\dr T\over \dr
t}){E\over T} - H {m^2\over ET} \right) f(p,z)^2 e^{\beta E+z}, \cr
}\eqno(8)$$
where $g$ is the neutrino spin factor.  Changing the variable from $t$ to
$x \equiv m/T$ and inserting (6-7) into (3), we obtain the equation for $z(x)$
of the form (2):
$$
{\dr z \over \dr x} =
(A(z) {x\over T}{\dr T\over \dr t})^{-1}
\left( - B(z) + S_{\rm I}(z) \right).
\eqno(9)$$
The function $B(z)$ on the R.H.S.\ of the equation (9) represents the
"free" part due to the expansion of the universe and the function
$S_{\rm I}(z)$ is the interaction term.  The remaining part of this section is
involved in writing (9) in a more explicit way.

Let us define the functions $J_\nr(x,z)$ as
$$
J_\nr(x,z) \equiv \int_1^\infty \dr y y^\nr (y^2-1)^{1\over 2}
{e^{xy}\over (1+e^{xy+z})^2}.
\eqno(10)$$
With help of these integrals we can write the functions $A(z)$ and $B(z)$
as
$$
\eqalign{
A(z) &= {g\over 2\pi^2}m^3e^z J_1(x,z)\cr
B(z) &= {g\over 2\pi^2}m^3e^z x \lbrace \left(H+{1\over T}{\dr T\over \dr t}
\right)J_2(x,z) - HJ_0(x,z) \rbrace .\cr
}\eqno(11)$$
The collision integral $S_{\rm I}(z)$ can always be reduced to
5-dimensions. We will choose our independent variables as the three
energies $E_1, E_2$
and $E_3$ and the two angles, $\theta$, the angle between the incoming
particles and $\phi$, the (acoplanarity) angle between the planes of the
incoming and outgoing particles.
With this choice of parameters one obtains, after elementary albeit
somewhat lengthy manipulations the result
$$
S_{\rm I}(z) = {m^4\over 512\pi^6}(e^{2z}-1)\sum_\nr \int {\cal D}\Phi
\int_0^{2\pi}\dr \phi \sum_{\rm spins}
{\mid {\cal M}_{\rm I}^{(\nr)}(u,v,t,\cos \theta ,\phi )\mid }^2,
\eqno(12)$$
where
$$\eqalign{
\int {\cal D}\Phi \equiv &\int_1^\infty \dr u \int_1^\infty \dr v
\int_{-1}^{1} \dr \cos \theta {\cP_u \cP_v \over
\kappa (u,v,\theta )}
\int_{t_-}^{t_+} \dr t \quad e^{xu+xv}\cr
&\times (e^{xu+z}-1)^{-1}(e^{xv+z}+1)^{-1}
(e^{xt}+1)^{-1}(e^{x(u+v-t)}+1)^{-1},\cr
}\eqno(13)$$
where $\cP_\alpha \equiv (\alpha^2-1)^{1/2}$ and
$\kappa (u,v,\theta ) \equiv \mid {\bf p}_1 + {\bf p}_2 \mid /m =
(\cP_u^2 +\cP_v^2 - 2\cP_u\cP_v\cos\theta)^{1/2}$
and we have used the scaled variables $u\equiv E_1/m$,
$v\equiv E_2/m$ and $t\equiv E_3/m$. The
$t$-integration limits are given by $t_\pm = {1\over 2}
(u+v \pm \kappa(u,v,t) \sqrt{1-4m_{\rm n}^2/s})$,
where $m_{\rm n}$ is the mass of the species n and
$s$ is the usual invariant
$s \equiv (p_1+p_2)^2 = 2m^2(uv+1-\cP_u \cP_v \cos\theta)$.
In some cases the matrix element $\cal M$ may be simple enough to allow
further integrations in (12), but due to the FD-distribution functions
a complete analytic evaluation is not possible even in the simplest
case of a constant matrix element.
Especially in the simple though not very realistic case of
$s$-dependent matrix element squared (12) can be integrated down
to two dimensions.

Finally, in order to fix the time temperature relation we will assume that
the universe expands adiabatically. Then
$$
{1\over T}{\dr T\over \dr t} =
- {H\over 1 + {1\over 3}{T\over h}{\dr h\over \dr T}},
\eqno(14)$$
where the function $h(T)$ is related to the entropy density of the
 interacting species $s$ by $s(T) \equiv {2\pi^2\over 45}h(T)T^3$.

Altogether we can now write the equation (9) into a more explicit form
$$
\eqalign{
{\dr z\over \dr x} = &- {J_0(x,z)\over J_1(x,z)} +
{1\over 3}{T\over h}{\dr h\over \dr T}
\left( {J_2(x,z) - J_0(x,z)\over J_1(x,z)}\right) \cr
&+\sinh z \phantom {l} \gamma (T)
{gxM_{\rm Pl}\over 128\pi^4mJ_1(x,z)} \cr
&\phantom{kat} \times \sum_\nr \int {\cal D}\Phi
\int_0^{2\pi} \dr \phi {1\over g^2}\sum_{\rm spin}
{\mid {\cal M}_{\rm I}^{(\nr)}(u,v,t,\theta,\phi) \mid }^2,\cr
}\eqno(15)$$
where $\gamma (T) \equiv  ({4\pi^3\over 45} g(T))^{-1/2}
\left(1 + {1\over 3}{T\over h}{\dr h\over \dr T}\right)$.
This is the main result of this section.
The equation (15) is actually simpler
than what it looks; it is only the matrix element that needs to be calculated
separately for each different case.  Of course (15) can only be solved
numerically, but there, even the multidimensional integrations are usually
rather easy and the equation itself is well behaving.

If $z(x)$ is known
one can obtain various thermodynamical quantities as
simple integrals over the momentum
$$
\alpha (x) \equiv {g \over 2\pi^2} \int_0^\infty \dr y y^2
{\bar \alpha (T,x)} (e^{(y^2+x^2)^{1/2}+z}+1)^{-1},
\eqno(16)$$
where e.g.\ ${\bar \alpha} = T^3$ and ${\bar \alpha} = (y^2+x^2)^{1/2}T^4$
for the number- and the energy densities respectively.
\jump
\centerline {\subsection 3. The Maxwell-Boltzmann limit}
\jump
We will now derive the Maxwell-Boltzmann limit of the equation (15).
In the MB limit the distribution function for neutrinos (5a) is
approximated by
$$
f(p,z)_{\rm MB} = e^{-\beta E -z}.
\eqno(17)$$
With this simplification the functions $J_\ir(x,z)$ can be expressed
in terms of the modified Bessel functions of the second kind:
$$
\eqalign{
J_\ir (x,z) &\rightarrow e^{-2z}{1\over x}K_{\rm i+1}(x)
\phantom {katti} ; \phantom{katti} i=0,1\cr
J_2 (x,z) &\rightarrow e^{-2z}{1\over x}\left( {3\over x^2}K_2(x) +
{1\over x}K_1(x) \right).\cr
}\eqno(18)$$
Secondly, the collision integral $S_{\rm I}(z)$ simplifies
considerably.  From (6) one obtains
$$
\eqalign{
S_{\rm I}(z)_{\rm MB} =
& (1-e^{-2z}) g^2 \int \prod_{\ir =1}^2
{\dr^3p_\ir \over (2\pi )^32E_\ir} e^{-\beta (E_1+E_2)} \cr
& \times \sum_\nr \int \prod_{\ir =3}^4
{\dr^3p_\ir \over (2\pi )^32E_\ir} (2\pi )^2
\delta^4(p_1+p_2-p_3-p_4){1\over g^2} \sum_{\rm spin}
{\mid {\cal M}_{\rm I}^{(\nr )}\mid }^2. \cr
}\eqno(19)$$
The second line in (19) represents a Lorentz invariant quantity,
namely the spin averaged cross section $\sigma^{(\nr )}(s)$ times the flux
$F\equiv 4((p_1\cdot p_2)^2-m^4)^{1/2} \equiv 4E_1E_2v_{\rm M\oo l}$, where
the flux-related velocity factor $v_{\rm M\oo l}$ is called the M\oo ller
velocity \refthree{9}{7}{5}.  Thus
$$
S_{\rm I}(z)_{\rm MB} = (1-e^{-2z})n_{\rm eq}^2(x) \sum_{\rm n}\vs ,
\eqno(20)$$
where $n_{\rm eq} \equiv g\int {\dr^3p \over (2\pi )^3} e^{-\beta E} =
g{Tm^2\over2\pi^2}K_2({m\over T})$ and the averaged cross section
$\vs$ is defined as
$$
\vs \equiv {4\pi^4\over m^4T^2K_2^2({m\over T})}
\int {\dr^3p_1 \over (2\pi )^3} \int {\dr^3p_2 \over (2\pi )^3}
e^{-\beta (E_1+E_2)} v_{\rm M\oo l}\sigma^{(\nr)}(s).
\eqno(21)$$
It is easy to reduce $\vs$ into an one-dimensional integral over $s$.  We
borrow the result of such an integration from the paper by Gondolo and Gelmini
\ref{5}
$$
\vs = {1\over 8m^4TK_2^2({m\over T})} \int_{4m^2}^\infty
\dr s \sqrt {s}(s-4m^2)K_1({\sqrt s\over T}) \sigma_{\rm CM}^{(\nr )}(s).
\eqno(22)$$
Using the equations (12), (18) and (20) the equation (9) can be written in the
MB-limit as
$$
{\dr z \over \dr x} = -{K_1(x)\over K_2(x)} +
{1\over x}{T\over h}{\dr h\over \dr T}
+\sinh z \phantom{l}
{g\over \pi^2}\gamma (T) mM_{\rm Pl}K_2(x)\sum_\nr \vs ,
\eqno(23)$$
where the function $\gamma (T)$ is as defined below the equation (15).
Of course, in the MB-limit we can write (22) directly in terms of the
number density $n$.  First note that the R.H.S\ of the equation (7) can
always be written as ${\dr n\over \dr t} + 3Hn$ and then use
$n \equiv n(T,z)_{\rm MB} = e^{-z}n_{\rm eq}(T)$ in the expression (20).
Then from (3), (7) and (20) one obtains the familiar result
$$
{\dr n\over \dr t} + 3Hn  = \sum_\nr \vs (n_{\rm eq}^2 - n^2).
\eqno(24)$$
\jump
\centerline {\subsection 4. The constant matrix element case}
\jump
We now wish to study quantitatively the size of the corrections arising from
the use of the FD-statistics.  To this end we consider the simplest case of a
constant matrix element interaction and
furthermore, we will neglect the entropy production, so that
${\dr h\over \dr T} \equiv 0$.
In order to mimic the strength of the standard
model weak interactions we choose to parametrize the matrix element as
${\cal M}_{\rm I} = \alpha G_{\rm F}^2m^4$, where $\alpha$ is a free
parameter. A value $\alpha \simeq 20$ would correspond to the interaction
strength of an ordinary neutrino in the early universe at few MeV
temperature.  Accordingly, we will fix $g(T) = 10.75$ corresponding to
$T \simeq {\cal O}(1-100)\rm MeV$.  We then obtain from (15) a simplified
equation
$$
{\dr z \over \dr x} = -{J_0(x,z)\over J_1(x,z)} - 3.08\times 10^{-4}\alpha
({m\over {\rm MeV}})^3 \sinh z {x\over J_1(x,z)} \int {\cal D}\Phi,
\eqno(25)$$
where, among other things, the trivial $\phi$-integration has been performed.
As an aside let us note that in this case, and in fact in the more general
case where $\mid {\cal M}_{\rm I}\mid ^2$ depends only on $s=(p_1 +p_2)^2$,
the integral over $\int {\cal D}\Phi$ can be reduced to a two-dimensional
one
$$
\eqalign{
\int {\cal D}\Phi {\mid {\cal M}_{\rm I} \mid }^2(s)
= 4\pi &\sum_\nr \sum_{\rm spin}
\int_{2x}^\infty \,dp \int _0^{(p^2-4x^2)^{1/2}} \,dq \, e^{-p}
{\mid {\cal M}_{\rm I} \mid }^2(s) \cr
&\times
\ln \left( {\cosh {1\over 4}(p+q)\over \cosh {1\over 4}(p-q)} \right)
\ln \left( {\cosh ({1\over 4}(p+qV) + {z\over 2}) \over
            \cosh ({1\over 4}(p-qV) + {z\over 2})} \right)
\cr}\eqno (26)$$
where $V=\sqrt {1-4m^2/s}$ and $s=T^2(p^2-q^2)=m^2(p^2-q^2)/x^2$.
The corresponding equation in the MB-limit is
$$
{\dr z \over \dr x} = -{K_1(x)\over K_2(x)} - 6.15\times 10^{-4}\alpha
({m\over {\rm MeV}})^3 {K_1^2(x)\over K_2(x)}  \sinh z.
\eqno(27)$$
We have solved numerically the equations (25) and (27).  In figure 1
we show our results for the evolution of the number density.
In all cases we have used the value $\alpha = 20$ for the
interaction strength.  Figure 1 shows that if the decoupling from
equilibrium occurs at $T \lsim {\cal O}(m)$, then the MB-approximation
overestimates the true value of the quantity in question,
the effect being largest in the region $T \gg m$.
In the last case however one can simply find the frozen number density
of particles in question without solving the kinetic equation
noting that it is equal to the equilibrium number density of massless
particles at the moment of decoupling.
In figure 2 we show the asymptotic value (for $T \simeq 0$)
of the scaled number density as a function of mass.
We also show the curves corresponding to an approximate analytic
solution of the equation (1) \reftwo{5}{10}
$$
f(0) \equiv {n\over s}  \simeq {H(T_{\rm f})\over s_{\rm f} \vs_{\rm f}}.
\eqno(28)$$
Here the index f refers to the freeze-out temperature, which is calculated
from the equation (assuming $\vs = \rm const.$)
$$
{m\over T_{\rm f}} \simeq \ln C - {1\over 2} \ln \ln C,
\eqno(29)$$
where $C$ is given by
$C \equiv ({4\pi^3\over 45}g(T))^{-1/2}
{g\over (2\pi )^{3/2}} \delta (\delta +2)
\alpha G_{\rm F}^2M_{\rm Pl}m^3/32\pi$.
For the free parameter $\delta$, we have used the generally accepted
"best fit" value $\delta (\delta +1) = 1$ \refthree{70}{5}{10}.
Here one can
see how the correction due to the FD-statistics dies away asymptotically
as the mass increases.  This reflects the fact that, while also
in this region
the MB-solution is initially inadequate, the particle ensemble remains in
equilibrium long enough for the asymptotics to be well described by the
MB-statistics. One should note also that expression (28) gives a very poor
approximation in the region of $m \lsim T_\dr$.
In particular when one is concerned with very
accurately known quantities such as restrictions derived from the
nucleosynthesis considerations, one should be careful to use at least the
accurate numerical solution of the equation (1) and in some cases the
numerical solution of the equation (15) with the correct statistics.
\jump
\centerline {\subsection 5. Conclusions and discussion}
\jump
We have derived an evolution equation for the pseudo-chemical potential
$z(T)$ in the early universe, that allows one to obtain various
thermodynamical quantities as a function the temperature,
as simple integrals over momentum, with the correct Fermi-Dirac statistics.
We have also obtained the reduction of this equation in the usual
Maxwell-Boltzmann approximation and compared this limiting case of our
result with
the literature.  We have solved the evolution equations numerically
for a simple toy model, where the interaction matrix element was assumed to
be a constant.  Our results confirm the expectation that the FD-corrections
are sizable in the region, where the decoupling from the thermal equilibrium
occurs at temperatures of the order of, or less than the neutrino mass.
Especially one should note that the use of some popular analytic
approximations for the relic
number density in this region is very ill advised.

The use of correct statistics could be important for example in the case of
a few MeV tau neutrino, which clearly would decouple while semirelativistic.
It has been noted elsewhere \ref{6} that such particles would strongly
influence the nucleosynthesis due to their potentially large energy density,
that would tend to increase the expansion rate of the universe. In these
studies the MB-approximation was used. As follows from our results the correct
statistics gives 5-10\% smaller results and correspondingly weaker bounds
on tau neutrino mass.

Let us finally mention that we have assumed here that both helicity states
are equally populated. It might not be true for the chiral interactions.
Especially this is the case for the standard model
interactions and therefore for the aforementioned tau-neutrino.
This problem together with calculation of Fermi corrections to the realistic
case of tau-neutrino freezing
will be pursued elsewhere \ref{8}.

\vfill\eject
\jump
\centerline {\subsection Acknowledgements}
\jump  A.D.\ wishes to thank the Center for Particle Astrophysics
at UC Berkeley where this work started and Physics Department
of University of Michigan where it was completed for the hospitality and
K.K.\ wishes to thank the Research Institute for Theoretical Physics,
University
of Helsinki, where part of this work was done for hospitality and
the Emil Aaltonen foundation and the Finnish Academy for financial support.
This research is supported by the DOE grant DE-AC02-83ER40105.
\vfill\eject
\centerline {\subsection References}
\jump
\refitem{1} {M.I.\ Vysotsky, A.D.\ Dolgov, and Ya.B.\ Zeldovich, JETP Letters
{\bf 26} (1977) 188,
B.W.\ Lee and S.\ Weinberg, Phys.\ Rev.\ Lett.\ {\bf 39}
(1977) 165.}
\refitem{70} {G.\ Steigman, Ann.\ Rev.\ Nucl.\ Part.\ Sci.\ {\bf 29} (1979)
313,
R.J.\ Sherrer and M.S.\ Turner, Phys.\ Rev.\ {\bf D33} (1986) 1585.}
\refitem{100} {K.\ Griest and D.\ Seckel, Nucl.\ Phys.\ {\bf B283} (1987) 681.}
\refitem{71} {Ya.B.\ Zeldovich, L.B.\ Okun, and S.B.\ Pikelner, Sov. Phys.
Uspekhi {\bf 8} (1965) 702.}
\refitem{2} {J.\ Bernstein, L.S.\ Brown and G.\ Feinberg, Phys.\ Rev.\
{\bf D32} (1992) 3261.}
\refitem{3} {K.\ Enqvist, K.\ Kainulainen and J.\ Maalampi, Nucl.\ Phys.\
{\bf B316} (1988) 456.}
\refitem{101} {D.E.\ Brown and L.J.\ Hall, Phys.\ Rev.\
{\bf D41} (1990) 1067,
see also reference \ref{10} and the references therein.}
\refitem{4} {M.\ Srednicki, R.\ Watkins and K.A.\ Olive, Nucl.\ Phys.\
{\bf B310} (1988) 693.}
\refitem{5} {P.\ Gondolo and G.\ Gelmini, Nucl.\ Phys.\ {\bf B360} (1992) 145.}
\refitem{6} {E.W.\ Kolb et.\ al.\ ,  Phys.\ Rev.\ Lett.\ {\bf 29} (1991) 533.}
\refitem{72} {A.D.\ Dolgov Sov.\ J.\ Nucl.\ Phys.\ {\bf 32} (1980) 831.}
\refitem{7} {S.R.\ de Groot, W.A.\ van Leeuwen and Ch.G.\ Weert, {\it
Relativistic kinetic theory} (North-Holland, Amsterdam, 1980)}
\refitem{8} {A.D.\ Dolgov and K.\ Kainulainen, in preparation.}
\refitem{9} {L.D.\ Landau and E.M.\ Lifshitz, {\it The classical theory of
fields} (Pergamon, Oxford 1975).}
\refitem{10} {E.W.\ Kolb and M.S.\ Turner, {\it The Early Universe}
(Addison-Wesley, Redwood city, CA, 1990)}
\printreference
\jump
\vfill\eject
\centerline {\subsection Figure Captions}
\jump
\noindent {\bf Figure 1}
The ratio of the number density of a particle of mass $m=5MeV$ and
$m=20MeV$ to the number density of massless particles $n/n(m=0)$
as a function of the
temperature $T$.  Solid lines correspond to the FD-statistics and the dashed
lines to the MB-statistics.  The dotted line shows the equilibrium number
density.
\jump
\noindent {\bf Figure 2}
The asymptotic limit ($T \simeq 0$) of the scaled number density
$n/n(m=0)$ as a function of the neutrino mass.
Solid line corresponds to the FD-statistics, dashed line to the
MB-statistics and dash-dotted line to the approximation (28).
\vfill\eject